\documentclass[11pt,a4paper]{article}
\usepackage[utf8]{inputenc}
\usepackage{amsmath,amssymb,amsfonts}
\usepackage{graphicx}
\usepackage[export]{adjustbox} 
\usepackage{booktabs}
\usepackage{multirow}
\usepackage{hyperref}
\usepackage{cite}
\usepackage{algorithm}
\usepackage{algorithmic}
\usepackage[margin=1in]{geometry}
\usepackage[numbers]{natbib}
\usepackage{times}

\title{Syllabic Agglutinative Tokenizations for Indonesian LLM:\\
A Study from ``Gasing Literacy Learning System''}

\author{%
  Hokky Situngkir\thanks{Bandung Fe Institute \& AI Research Center IT Del, \texttt{hokky.situngkir@del.ac.id}}
  \and
  Andhika Bernard Lumbantobing\thanks{Bandung Fe Institute, \texttt{nad@compsoc.bandungfe.net}}
  \and
  Yohanes Surya\thanks{AI Research Center IT Del, \texttt{yohanessurya@del.ac.id}}
}

\date{Version: 2025-08-04}

\begin{document}

\maketitle

\begin{abstract}
This paper presents a novel syllable-based tokenization approach for Indonesian large language models, inspired by the Gasing Literacy Learning System's pedagogical methodology. Drawing on information-theoretic principles, we develop a tokenization framework that segments Indonesian text at syllable boundaries before applying byte-pair encoding, creating a vocabulary that aligns with the language's morphophonological structure. Our approach first identifies high-frequency syllables through rule-based segmentation, then constructs a compact vocabulary of 3,500 tokens that preserves meaningful linguistic units while maintaining coverage through character-level fallback. Empirical evaluation on Indonesian Wikipedia and folklore corpora from Indonesian Culture Digital Library (PDBI) demonstrates substantial improvements over conventional tokenization methods: the syllable-based approach achieves Rényi efficiency of 0.74 compared to 0.50-0.64 for pretrained multilingual tokenizers, while maintaining higher average token lengths (3.67 characters versus 2.72 for GPT-2) despite using a vocabulary an order of magnitude smaller. These gains emerge from the method's ability to internalize character-level dependencies within syllable units, reducing the computational burden on language models while respecting Indonesian's agglutinative morphology. We call the LLM built upon this principle, TOBA LLM (Tokenisasi Optimum Berbasis Aglutinasi), the convergence of human literacy pedagogy with computational optimization principles offers a promising paradigm for developing linguistically-informed tokenization strategies, particularly for morphologically rich and underrepresented languages in natural language processing.
\end{abstract}


\section{Introduction}

The rapid development of large language models (LLMs) has had a substantial impact on many languages, including Indonesian. However, most open-source models and fine-tuned variants that support Indonesian still employ tokenization schemes trained in multilingual settings. This approach enlarges the vocabulary size and reduces computational efficiency, particularly when the model is used for a single language. Indonesian, moreover, is characterized by its highly productive system of affixation: attaching affixes to root words can significantly alter their meaning. The language also absorbs a wide range of loanwords from regional tongues, encompassing everyday vocabulary as well as technical terms, cultural expressions, and distinctive morphological forms. Tokenization methods that overlook these local characteristics risk misidentifying meaningful units—especially when those tokens are absent from the pretrained model's vocabulary.

The ``Gasing Literacy Learning System'' is an instructional approach to Indonesian that integrates reading, writing, and pronunciation while explicitly attending to local linguistic features—beginning with the construction of words from syllables and progressing to fully contextualized sentences. Developed by Indonesian physicist Yohanes Surya, the method offers an alternative language-teaching pathway for students who speak a wide array of regional languages. Instruction proceeds from identifying syllabic building blocks, to accurate articulation, and ultimately to very rapid reading and writing proficiency. The system draws on the fact that Indonesian shares key morphological and transliteration patterns with the hundreds of local languages spoken across the archipelago.

GASING is an acronym for \textit{Gampang, Asyik, Menyenangkan} (``Easy, Fun, and Enjoyable''). Gasing Literacy represents the foundational literacy arm of the broader GASING educational program, which has been used with thousands of Indonesian children to strengthen reasoning, language, mathematics, and science competencies\footnote{GASING Academy, URL: \url{https://gasingacademy.org/}}. Within the same paradigm, this tokenization model and language-literacy framework is ``introduced'' to a large language model in order to assess specific potentials, relative to the tokenization schemes currently in widespread use among mainstream LLMs.

In this study, tokenization is designed and implemented on an Indonesian-specific syllabic agglutination basis. Syllable-agglutinative tokenization is selected because it potentially offers several advantages:

\begin{itemize}
    \item \textbf{Smaller vocabulary:} Representing the corpus with fewer tokens lowers the overall vocabulary size, directly shrinking the embedding and output layers of an LLM and making memory and cache use more efficient.
    \item \textbf{Data-efficient generalization:} Because syllable tokens follow well-defined linguistic rules, the model can generalize more quickly with less training data.
    \item \textbf{Faster, more stable training:} The natural syllabic structure of Indonesian speeds up convergence and improves training stability.
    \item \textbf{Greater sensitivity to local languages:} Capturing morphological and phonological patterns at the syllable level makes the tokenizer more responsive to regional language varieties.
\end{itemize}

The tokenizer is built with downstream Indonesian applications in mind and draws on information-theoretic principles. To assess its performance, we conduct a series of experiments and compare the syllabic-agglutination-based approach with popular algorithms—Byte-Pair Encoding (BPE), WordPiece, and the pretrained GPT-2 and BERT tokenizers. Evaluation employs intrinsic metrics (token distribution, compression efficiency, and vocabulary size) and extrinsic metrics obtained by measuring GPT-2 \cite{radford2019language} performance on a range of natural-language-processing tasks.

\section{Methods}

\subsection{Unigram Model}

For a stationary and ergodic source $P$, satisfying $p(X_1, \ldots, X_n) = p(X_{1+k}, \ldots, X_{n+k}) \forall k$ and generating a symbol sequence, $X_1, X_2, \ldots, X_n \sim P$, the entropy rate per symbol is defined as

\begin{equation}
H(P) = \lim_{n \to \infty} \frac{1}{n} H(X_1, X_2, \ldots, X_n)
\end{equation}

By Shannon's source-coding theorem \cite{shannon1948mathematical, cover2006elements}, this limit is the optimal lossless compression rate for $P$. If a model $Q$ is used to approximate the distribution $P$, the cross-entropy loss of $Q$ with respect to $P$ is

\begin{equation}
L_n(Q) = H(P, Q) = E_P\left(\log \frac{1}{Q}\right)
\end{equation}

The quantity $H(P, Q) - H(P) = E_P\left(\log \frac{P}{Q}\right) = D_{KL}(P||Q) \geq 0$ is the relative entropy (Kullback-Leibler divergence) between $P$ and $Q$, which measures the inefficiency incurred when one assumes the model $Q$ while the true distribution is $P$. Relative entropy is always non-negative and equals zero only when $Q = P$. Accordingly, $H(P)$ constitutes the theoretical lower bound on the cross-entropy achievable by any model $Q$ over sequences $X \sim P$.

\begin{equation}
\min_Q L_n(Q) = H(P)
\end{equation}

A simple approximation to $P$ is the unigram symbol model $Q \subseteq Q_{1-gram}$, which assumes that the symbols in the sequence $X \sim P$ are independent, so that

\begin{equation}
q(X_1, \ldots, X_n) = \prod_{i=1}^n q(X_i)
\end{equation}

If the marginal symbol distribution $\lim_{n \to \infty} \frac{1}{n} \sum p(X_i = x) = \pi(x)$, is known, we obtain the cross-entropy for $Q$:

\begin{align}
L_n(Q) &= \lim_{n \to \infty} \frac{1}{n} E_P\left(\log \frac{1}{q(X_1, \ldots, X_n)}\right) \\
&= \lim_{n \to \infty} \frac{1}{n} \sum_{i=1}^n E_P(-\log q(X_i)) \\
&= \frac{1}{n} \sum_{i=1}^n E_\pi(-\log q(x)) = H(\pi, Q)
\end{align}

and $L_n(Q) = H(\pi, Q) \geq H(\pi)$, with equality $L_n(Q) = H(\pi)$ attained only when $q = \pi$. In empirical work, the distribution $\pi$ is usually estimated by counting symbol frequencies in the corpus. Even so, the value $H(\pi)$ will approach $H(P)$ only when $P(X_1, \ldots X_n) \approx P(X_1)P(X_2)\ldots P(X_n)$. For natural processes such as human language, this independence assumption is obviously problematic. A direct consequence of ignoring inter-symbol dependencies is a rise in the uncertainty of sequences generated by $P$ when they are modeled with $Q$; this can be demonstrated by decomposing the cross-entropy of the unigram model $Q$:

\begin{align}
L_n(Q) &= \lim_{n \to \infty} \frac{1}{n} E_P\left(\log \frac{1}{Q}\right) \\
&= \lim_{n \to \infty} \frac{1}{n} E_P\left(\log \frac{1}{P(X)}\right) + \lim_{n \to \infty} \frac{1}{n} E_P\left(\log \frac{P(X)}{Q(X)}\right) \\
&= H(P) + \lim_{n \to \infty} \frac{1}{n} E_P\left(\log \frac{P(X_1,\ldots,X_n)}{P(X_1)\ldots P(X_n)}\right)
\end{align}

The mutual information $I(X_1; \ldots; X_n) = E_P\left(\log \frac{P(X_1,\ldots,X_n)}{P(X_1)\ldots P(X_n)}\right)$ in eq. (8) is always non-negative and equals zero only when all symbols generated by $P$ are independent. In other words, the gap between the model loss $L_n(Q)$ and the true entropy rate $H(P)$ quantifies the amount of inter-symbol dependency in sequences $X \sim P$ that the unigram model $Q$ fails to capture.

In downstream text-based language-modelling tasks—especially text-generation objectives such as causal language modelling—treating single characters as the prediction unit places a heavy burden on training more complex predictors. The model must first master the dependencies among characters before it can capture higher-level linguistic regularities such as semantics or syntax \cite{libovicky2021why, itzhak2021models}. Behavioural studies of Transformer models \cite{vaswani2017attention} on symbol sequences produced by simple Markov processes likewise show that, even with self-attention, the model needs substantially more time—or may fail outright—to learn the true sequence distribution when its prediction targets are single-character symbols \cite{makkuva2024attention, rajaraman2025toward}. In such situations the model ends up predicting symbols from their marginal distribution, effectively behaving no better than a unigram model.

Although a unigram model relies on very simple assumptions, its loss can be pushed approaching $H(P)$ by internalizing local dependencies in the sequence $X \sim P$ through segmentation (cf. \cite{shannon1951prediction}). Here, the model $Q_{seg}$ operates not on individual characters but on segments—clusters of symbols that frequently co-occur and are treated as a single prediction unit. For example, if $P(xx') \gg P(x)P(x')$, elevating $xx'$ to a prediction unit makes the resulting unit distribution more nearly independent and can lower the unigram loss.

For a segment-based unigram model $Q_{seg}$ and a segmentation $T: X^* \to T^*$ that maps a sequence of single symbols $x_1, x_2, \ldots, x_n$ to a sequence of segments $t_1, t_2, \ldots, t_k$ with $t_i \in T$ and $\sum_{i=1}^k |t_i| = n$, the per-symbol entropy rate can be expressed as follows:

\begin{equation}
L_n(Q \circ T) = \frac{H(Q_{seg})}{E(L)}
\end{equation}

with $H(Q_{seg}) = E_P\left(-\sum_{i=1}^k \log q(t_i)\right)$ denote the cross-entropy of the segment-unigram model after the symbol sequence has been segmented by $T$ and $q: T \to [0,1]$ is the unigram distribution over segments and $\sum_{t \in T} q(t) = 1$. $L_n(Q \circ T)$ measures the cross-entropy of the segment-based unigram model per input character. Here $E(L) = \sum_{i=1}^k q(t_i) \cdot |t_i|$. Low $H(Q_{seg})$ implies that the segments returned by $T$ supply a distribution that predicts sequences from the source $P$ more accurately, while high $E(L)$ means that longer segments occur frequently, so the original symbol sequence can be expressed more compactly in terms of those segments. Together, the pair $H(Q_{seg})$ and $E(L)$ offers concrete criteria for evaluating and selecting a segmentation scheme: the smaller the value of $L_n(Q \circ T)$ (i.e.: the closer it approaches the theoretical lower bound $H(P)$ the more optimal the segmentation.

\subsection{Tokenization}

Tokenization is a standard preprocessing step in text-based large language models (LLMs). Neural-network text processing begins by segmenting raw text into a sequence of units called tokens. Formally, the tokenization function $t: \Sigma^* \to D \subseteq \Delta^*$ maps a character string $\sigma = \langle \sigma_1, \ldots, \sigma_n \rangle$ over the alphabet $\Sigma$ to a token sequence $D = t(\Sigma^*)$ over the token alphabet $\Delta$. An LLM works with a fixed vocabulary $\Delta$; each token $\delta \in \Delta$ is given a numerical representation (embedding) that is learned during training.

Tokenization methods can be grouped into three broad categories:

\begin{enumerate}
    \item \textbf{Word-level.} Entire words are treated as tokens, so the resulting sequences are short. The drawback is a very large vocabulary—needed to cover every word type—and poor coverage of out-of-vocabulary text.
    
    \item \textbf{Character-level.} Single characters form the tokens, giving excellent robustness to spelling variation and unseen words. As discussed earlier, however, predicting one character at a time burdens the model with learning low-level dependencies and lengthens the token sequence, increasing context-window requirements and computational cost.
    
    \item \textbf{Subword-level.} Schemes such as BPE or SentencePiece split text into units larger than a character but smaller than a whole word, aiming to capture the advantages of both extremes \cite{kudo2018subword, sennrich2016neural}.
\end{enumerate}

A central design goal is to ensure that the resulting token-frequency distribution is balanced. Tokens that occur far too frequently often carry little information, while those that occur too rarely receive scant exposure during training, preventing the model from learning reliable representations.

Given a unigram token distribution $p_\Delta$, we can compute the Rényi entropy:

\begin{equation}
H_\alpha(p_\Delta) = \lim_{\alpha' \to \alpha} \frac{1}{1-\alpha'} \log \left(\sum_{\delta \in \Delta} p_\Delta(\delta)^{\alpha'}\right)
\end{equation}

When $\alpha = 1$, $H_1(p_\Delta) = H(W_\Delta) = -\sum_{\delta \in \Delta} p_\Delta(\delta) \log p_\Delta(\delta)$ reduces to Shannon entropy, which gives the theoretical lower bound on the optimal code length for the tokens. In general, $H_\alpha$ penalises token distributions that contain extremely high- or low-frequency items; therefore, a higher entropy value is desirable, as it indicates a more balanced token distribution. To compare tokenizations with different vocabulary sizes, we can define the Rényi efficiency as:

\begin{equation}
\eta_\alpha = \frac{H_\alpha(p_\Delta)}{\log(|\Delta|)}
\end{equation}

The efficiency $\eta_\alpha \in [0,1]$ is the ratio between the optimal code length (with the lower bound given by entropy $H_\alpha$) and the expected code length assuming a uniform token distribution. A higher efficiency indicates a closer data compression rate to the optimal and a more balanced token distribution. Empirical studies of downstream tasks \cite{zouhar2023tokenization} further show that Rényi entropy at $\alpha = 2.5$ correlates positively with LLM performance, as measured by the BLEU translation-quality metric.

\subsection{Syllabic Agglutinative Tokenizer Design}

\subsubsection{Rule-based Syllable Segmentation}

To build a tokenizer whose vocabulary is well-suited to Indonesian, the corpus is first pre-processed with a syllable-segmentation rule $T: X^* \to T^*$ where $X$ is the set of character symbols and $T$ is the set of syllables.

This segmentation step yields the initial alphabet $\Sigma$, which serves as the basis for token construction. Applying $T$ to the raw corpus $\{x_1, \ldots, x_M\} \subset X^*$ gives segmented sequences from which we estimate the unigram syllable distribution:

\begin{equation}
p_T(t) = \sum_{t \in T^*} p_{T^*}(t) \frac{count(t,t)}{|t|}
\end{equation}

with $t = T(x)$. Syllables that occur with high frequency typically capture strong local character-level dependencies in natural Indonesian text. Conversely, very low-frequency syllables tend to be uncommon in ordinary writing; they often correspond to loanwords (e.g., from English or Latin) lacking established local variants, or to technical terms and rare named entities.

High-frequency syllables are therefore retained as elements of the alphabet $\Sigma$, whereas low-frequency syllables are discarded to avoid under-exposure during language-model training.

To mitigate the inevitable fraction of out-of-vocabulary text, basic symbols—such as raw Unicode characters—are still included in $\Sigma$, mirroring the byte-level strategy widely used in Byte-Pair Encoding (BPE).

Given $T$ and $\Sigma$, we define the practical segmentation

\begin{equation}
T_\Sigma = (f_\Sigma \circ T)
\end{equation}

where $f_\Sigma$ maps any syllable $t \notin \Sigma$ to its constituent single-character $t_1, \ldots, t_k \in \Sigma$. This segmentation $T_\Sigma$ is subsequently employed in the syllable-based tokenization.

\subsubsection{Syllable-Alphabet-Based Tokenization}

The tokenizer is designed in line with information-compression principles and implemented with the Byte-Pair Encoding (BPE) algorithm \cite{gage1994new, zouhar2024formal}. Using the alphabet set $\Sigma$ obtained from the initial syllable-segmentation procedure, we can define the set of minimal merge pairs $\Upsilon_\Sigma$ on $\Sigma$ that satisfies:

\begin{itemize}
    \item $\sigma \in \Sigma \Rightarrow \sigma \in \Upsilon_\Sigma$
    \item $\mu', \mu'' \in \Upsilon_\Sigma \Rightarrow [\mu', \mu''] \in \Upsilon_\Sigma$
\end{itemize}

A merge sequence $\mu = \langle \mu_1, \ldots, \mu_N \rangle \in \Upsilon_\Sigma^*$ is valid when, for every $\mu_n = [\mu', \mu'']$ and for each $\mu \in \{\mu', \mu''\}$, we have either $\mu = \mu_k$ with $k < n$ or $\mu \in \Sigma$. In other words, the pair merged at step $n$ must be composed of units that already exist in the alphabet $\Sigma$ or that were created by an earlier merge.

For any text $t \in \Sigma^*$, the BPE algorithm represents that text by iteratively applying segment merges according to the sequence $\mu$. Training a tokenizer with BPE is essentially a greedy search for the optimal merge sequence $\mu^*$ that maximises compression utility as follows:

\begin{equation}
\kappa_t(\mu) = |t| - |m_\mu(t)|
\end{equation}

where $|t|$ denotes the length of the text string and $|m_\mu(t)|$ is the length of the sequence obtained after applying the merge sequence $\mu$ to $t$. The BPE training procedure is run on the corpus $\{t_1, \ldots, t_M\} \subset \Sigma^*$ until the desired vocabulary size is reached.

With the syllable alphabet $\Sigma$, the practical segmentation $T_\Sigma$, and the learned merge sequence $\mu$, the tokenization function $e_\Sigma: X^* \to \Upsilon_\Sigma^*$ can be written as

\begin{equation}
e_\Sigma = (m_\mu \circ T_\Sigma)
\end{equation}

\section{Training}

\subsection{Segmentation Evaluation}

We conducted a characterization and comparison of the practical segmentation $T_\Sigma$, which is based on Indonesian syllable rules, with segmentation using Byte-Pair Encoding (BPE) tokenization. In practice, constructing the merge sequence during BPE training requires specifying the desired vocabulary size. To ensure a fair comparison, we selected BPE vocabulary sizes equivalent to the number of syllable alphabets resulting from the selection of high-frequency syllables.

Both segmentation methods were trained on the same corpus, namely the Indonesian Wikipedia corpus. This corpus consists of informative articles written in Indonesian covering various domains of knowledge, including science and technology, history, culture, and politics. For syllable-based segmentation, we applied a top-k selection criterion that choose the most frequent syllables, using $k_1 = 500$ and $k_2 = 1,000$ serving as the base vocabulary sizes. Single characters were added to address out-of-vocabulary (OOV) issues, resulting in vocabulary sizes of $|V_1| = 680$ and $|V_2| = 1,166$. These sizes were then used as reference points for setting the target vocabulary size in BPE tokenization.

Evaluation was conducted using two primary metrics on the training corpus: the unigram segment model loss and the average number of characters per segment produced by each segmentation method. Based on these metrics, we calculated bits per character (BPC) to measure how effectively each segmentation captures character-level dependencies within the corpus. The BPC of character-based unigram model was used as a baseline for comparison. In addition to evaluation on the training corpus, segmentation was also tested on the folklore corpus, which consists of folk stories collected through community participation in the Perpustakaan Digital Budaya Indonesia (PDBI) \cite{zawani2008platform}. Since this corpus was not involved in vocabulary training, it provides a reliable measure of the generalization capabilities of each segmentation.

Table \ref{tab:segmentation_eval} presents the evaluation results for the four segmentation methods. Both syllable-based and BPE-based segmentation achieve lower bits per character (BPC) compared to the character-level unigram loss. This indicates that both approaches are effective in capturing character-level dependencies within the corpus. Furthermore, we observe that increasing the vocabulary size leads to greater uncertainty at the segment level, as evidenced by the increase in unigram segment loss. However, this is accompanied by a decrease in BPC, which can be attributed to the increase in the average character length per segment as the vocabulary expands.

\begin{table}[h]
\centering
\caption{Segmentation Evaluation using the Unigram Model}
\label{tab:segmentation_eval}
\begin{tabular}{lcccccccc}
\toprule
\multirow{3}{*}{Segmentation} & \multirow{3}{*}{\begin{tabular}[c]{@{}c@{}}Vocabulary\\ size\end{tabular}} & \multicolumn{6}{c}{Unigram Model} \\
\cmidrule{3-8}
& & \multicolumn{2}{c}{Bits per Segment} & \multicolumn{2}{c}{Characters per Segment} & \multicolumn{2}{c}{Bits per Character} \\
& & WikiID & PDBI & WikiID & PDBI & WikiID & PDBI \\
\midrule
Character-based & - & - & - & - & - & 2.88 & 2.75 \\
\midrule
Syllable-based & \multirow{2}{*}{680} & 5.27 & 5.10 & 1.93 & 2.06 & 2.73 & 2.47 \\
BPE & & 5.72 & 5.43 & 2.38 & 2.43 & 2.40 & 2.24 \\
\midrule
Syllable-based & \multirow{2}{*}{1,166} & 5.69 & 5.40 & 2.17 & 2.28 & 2.62 & 2.37 \\
BPE & & 6.17 & 5.82 & 2.73 & 2.76 & 2.26 & 2.11 \\
\bottomrule
\end{tabular}
\end{table}

The underlying mechanisms driving the reduction in BPC differ between syllable-based and BPE-based segmentation. Syllable-based segmentation expands the vocabulary by including more high-frequency syllables. As a result, sequential characters identified as syllables are grouped into single segments, thereby reducing the number of character-based segments. On the other hand, BPE segmentation lowers BPC through the increasing number of merge sequences applied as the vocabulary grows. BPE iteratively merges frequently co-occurring pairs of characters or segments, forming new segments composed of longer character sequences.

While syllable-based segmentation yields lower unigram loss at the segment level, its resulting BPC are higher compared to BPE. This is due to BPE's ability to generate longer segments in terms of character length, aligning with the algorithm's objective of maximizing compression utility. Syllable-based segmentation, however, is inherently limited in segment length due to the natural structure of Indonesian syllables, which typically consist of only two to three letters. This restricts the potential for compression, as segment length cannot increase significantly even when the vocabulary is expanded. Moreover, the occurrence of rare syllables in the corpus that are not part of the predefined syllable alphabet causes syllable-based segmentation to fall back on splitting them into individual characters, further reducing compression utility.

\subsection{Intrinsic Evaluation of Tokenization}

Intrinsic evaluation of tokenization performance was conducted using two primary metrics: the average number of characters per token and token distribution efficiency as measured by Rényi efficiency. The syllable-alphabet-based tokenization was trained on the Indonesian Wikipedia corpus to determine the syllable distribution. The base syllables forming the alphabet were selected through top-k selection with $k = 1,500$, supplemented by the inclusion of individual Unicode characters in the alphabet. The BPE algorithm was then applied to expand the vocabulary until reaching a final size of 3,500 tokens. For comparison, a plain BPE tokenization was also developed by learning the vocabulary from the byte level without incorporating syllables as the base alphabet. Both tokenization methods were trained on the same corpus and with an equivalent vocabulary size.

To broaden the scope of evaluation, the performance of the two tokenization methods was also compared with the tokenization schemes of two widely adopted pre-trained models: GPT-2 and BERT. GPT-2 tokenization was developed using the byte-level BPE algorithm and has a vocabulary size of 50,257 tokens, while BERT tokenizer was trained using the WordPiece algorithm with a vocabulary size of 30,522 tokens. Evaluation across all tokenization methods was conducted using two corpora: the Indonesian Wikipedia training corpus and the folklore corpus from the Perpustakaan Digital Budaya Indonesia (PDBI).

The evaluation results are summarized in Table \ref{tab:tokenization_eval}. Despite having smaller vocabulary sizes, the tokenizers trained specifically on Indonesian-language corpora—both the syllable-alphabet-based and the plain BPE—achieve higher average character lengths per token than the pre-trained tokenizations. This indicates that these tokenizations generate more compact text representations with a higher frequency of longer tokens. In contrast, the pre-trained tokenizations of GPT-2 and BERT were trained on large, multilingual corpora that are not language-specific and are disproportionately dominated by English \cite{petrov2023language}. Although these pre-trained tokenizations possess significantly larger vocabularies, only a small fraction of their tokens are relevant or naturally occur in Indonesian language contexts. In other words, the pre-trained tokenizations are less optimal in capturing the linguistic structure specific to Indonesian, resulting in lower efficiency in text representation at the token level.

\begin{table}[h]
\centering
\caption{Tokenization Evaluation based on Average Characters per Token and Rényi Efficiency}
\label{tab:tokenization_eval}
\begin{tabular}{lcccccc}
\toprule
\multirow{2}{*}{Tokenization} & \multirow{2}{*}{\begin{tabular}[c]{@{}c@{}}Vocabulary\\ size\end{tabular}} & \multicolumn{2}{c}{Average Characters per Token} & \multicolumn{2}{c}{Rényi Efficiency ($\alpha = 2.5$)} \\
& & Wiki & PDBI & Wiki & PDBI \\
\midrule
Plain BPE & 3,500 & 3.55 & 3.41 & 0.64 & 0.62 \\
Syllable-based BPE & 3,500 & 3.67 & 3.44 & 0.74 & 0.68 \\
Pre-Trained GPT2 & 50,257 & 2.72 & 2.50 & 0.50 & 0.53 \\
Pre-Trained BERT & 30,522 & 3.48 & 3.32 & 0.49 & 0.54 \\
\bottomrule
\end{tabular}
\end{table}

Although both use the same token merging algorithm, i.e., BPE, the syllable-alphabet-based tokenization yields a higher average token length compared to plain BPE tokenization. This is due to the syllable-based approach having already internalized frequently co-occurring characters into its base alphabet units, aligning more with natural and functional linguistic boundaries, such as affixes. Given the same vocabulary size, the syllable-based is faster and more effective at capturing higher-level dependency patterns in text, particularly at the syllabic level. Plain BPE tokenization requires more merge steps to learn these higher-level dependencies, as it must construct them incrementally from individual characters. Moreover, this process often results in intermediate tokens that rarely occur independently, thereby expanding the ``unused token zone'' within the vocabulary \cite{bostrom2020byte}. This reduces the economic utility of plain BPE, especially in scenarios where vocabulary budget is constrained.

The evaluation results also show that the syllable-based tokenization achieves the highest Rényi efficiency on both the training and test corpora. This indicates that the resulting token distribution is more balanced, offering an advantage for training more complex predictive models by facilitating more uniform token representation learning.

\section{Discussions}

\subsection{Implications of Syllable-based Tokenization Performance}

The empirical results demonstrate that syllable-based tokenization achieves markedly superior performance across multiple evaluation metrics compared to conventional approaches. The Rényi efficiency values of 0.74 (Indonesian Wikipedia corpus) and 0.68 (folklore PDBI corpus) at $\alpha = 2.5$ represent substantial improvements over both pretrained multilingual tokenizers and plain BPE implementations. These efficiency gains directly correlate with more balanced token distributions, which facilitate uniform representation learning across the vocabulary space.

The observed compression characteristics warrant particular attention. Despite employing a vocabulary of only 3,500 tokens—an order of magnitude smaller than standard multilingual models—the syllable-based approach achieves average token lengths of 3.67 characters (Indonesian Wikipedia corpus) and 3.44 characters (folklore PDBI corpus). This compression efficiency emerges from the fundamental alignment between tokenization boundaries and Indonesian morphophonological structure, where syllables constitute natural units of phonological organization.

From an information-theoretic perspective, the superior performance can be attributed to the internalization of intra-syllable character dependencies within the base tokenization units. The cross-entropy decomposition reveals that by capturing these local dependencies at the syllable level, the burden on subsequent model layers to learn character-level patterns is substantially reduced. This architectural advantage becomes particularly pronounced in Indonesian, where syllable boundaries frequently coincide with morphological boundaries, enabling more effective learning of compositional semantics.

\subsection{Computational and Implementation Considerations}

The computational advantages of syllable-based tokenization extend beyond compression metrics. The deterministic nature of syllable segmentation yields $O(n)$ time complexity for text processing, compared to the potentially quadratic complexity of unrestricted BPE tokenization. This efficiency gain becomes increasingly significant for large-scale corpus processing and real-time inference applications.

The reduced vocabulary size directly impacts model architecture parameters. The embedding layer size reduction from typical values of 30,000-50,000 to 3,500 tokens translates to proportional reductions in memory requirements and parameter counts. For deployment scenarios with constrained computational resources—particularly relevant for Indonesian language technology adoption—these efficiency gains enable broader accessibility of language model capabilities.

\subsection{Linguistic and Cultural Alignment}

The success of syllable-based tokenization illuminates broader principles regarding the intersection of linguistic structure and computational representation. Indonesian's agglutinative morphology, characterized by systematic affixation patterns, benefits substantially from tokenization that preserves these meaningful units \cite{situngkir}. The approach naturally segments complex words while maintaining morphological transparency, facilitating more effective learning of grammatical patterns and semantic compositionality.

The connection to the Gasing Literacy Learning System introduces compelling parallels between human language acquisition and machine tokenization strategies. The pedagogical emphasis on syllable-level processing in human literacy education appears to translate effectively to computational contexts, suggesting fundamental cognitive and information-processing principles that transcend the human-machine distinction.

\section{Implementation as Language Model System}

We built an implementation of a language model system by integrating explicit linguistic knowledge-specifically, the syllable rules of Indonesian - as a preliminary stage prior to the application of statistical methods. We call this type of LLM tokenization: Tokenisasi Optimum Berbasis Aglutinasi (TOBA-LM), an LLM basic model employing form of optimum tokenizations based on natural linguistic agglutination.

The process proceeds through two main phases. First is the Rule-Based Syllable Segmentation phase (Initial Alphabet Construction), which leverages various Indonesian text corpora. Input texts are first processed by a deterministic, rule-based segmentation function that splits the text at syllable boundaries according to Indonesian orthographic conventions. From these segmentations, high-frequency syllable units are identified and used to form an initial “syllable alphabet,” reflecting local dependencies among character sequences in the text. To ensure coverage and robustness, basic Unicode code points are also explicitly included in this alphabet. Whenever a syllable is encountered that does not appear in the initial alphabet, the system automatically falls back to decomposing it into its constituent individual characters.

The second phase is the Alphabet-Based Unit Merging process, wherein the rule-segmented text - now represented as a sequence of syllable units and/or individual characters—is further processed by a merging algorithm. This algorithm iteratively combines adjacent units (including fallback characters) drawn from the initial alphabet to form longer, more complex tokens. The outcome of this process is a final vocabulary and a set of learned merge rules that emerge in a data-driven manner from the observed combinations of syllable units. By internalizing character relationships at the syllable level from the tokenization stage onward, the system allows the language model (LLM) to focus its learning capacity on higher-order linguistic patterns, such as semantics and syntactic structure. The initial alphabet, enriched with linguistically meaningful syllable units, functions as a form of linguistic priming, improving merging efficiency and facilitating the acquisition of semantically and morphologically cohesive units.

\begin{figure}[htbp]
  \centering
  \includegraphics[width=1\textwidth]{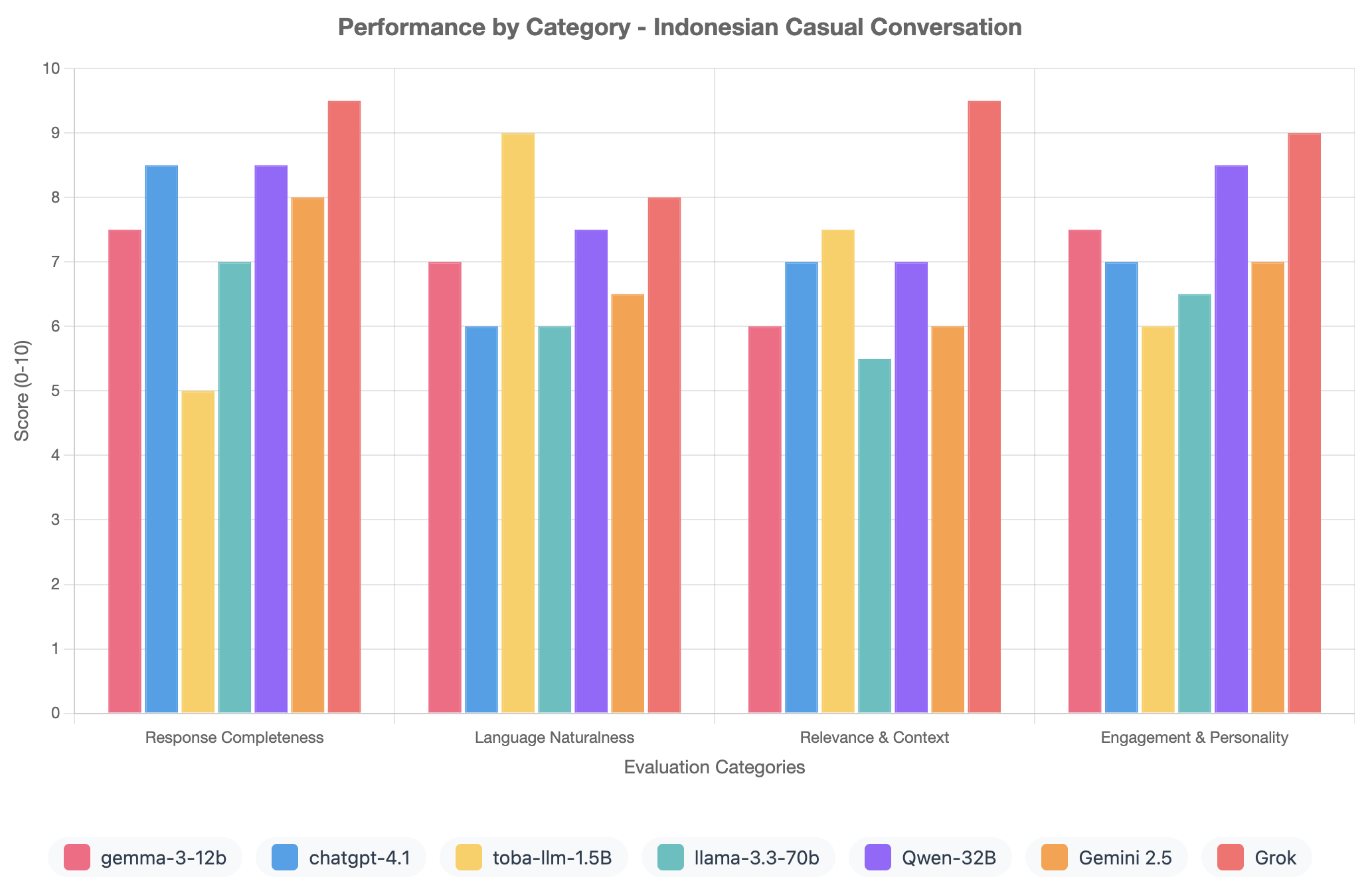}
  \caption{Comparison of TOBA-LM with other popular LM models chatbots}
  \label{fig:bm-gambar}
\end{figure}

On the basis of training with this system, we designed an inference protocol to conduct casual conversational experiments with TOBA-LM, comparing its prompt-response style against those of widely used online LLMs—such as Grok, ChatGPT 4.1, and Gemini 2.5—as well as offline models including Qwen 32B, Llama 3.3 70B, and Gemma 3 12B. TOBA-LM itself is a relatively small model, comprising 1.5 billion parameters, and is evaluated on casual Indonesian dialogue across both formal and informal, locally contextualized themes.

We employed Anthropic’s Claude Opus 4 to rate each model’s responses across four categories:
\begin{enumerate}
\item \textbf{Response Completeness:} How informative and conversation-stimulating the responses are, including depth of information and encouragement for follow-up discussion.
\item \textbf{Indonesian Language Naturalness:} The use of casual, natural, and colloquial Indonesian expressions that feel authentic in daily conversation.
\item \textbf{Relevance \& Context:} The ability to answer questions appropriately with local Indonesian context and cultural awareness.
\item \textbf{Engagement \& Personality:} Friendliness, interactivity, and the capacity to encourage continued conversation through engaging personality traits.
\end{enumerate}

The comparison showed that TOBA-LM excels in informal language usage—including terms that would be considered non-standard in formal Indonesian—and achieves higher naturalness scores in everyday Indonesian compared to all other tested LLMs. It also performs well in relevance and context, owing to its incorporation of local linguistic patterns, although Grok attained the highest scores in this category. Conversely, TOBA-LM lags behind in response completeness and conversational flow (engagement), which can be attributed to its more limited training knowledge base, given its relatively small parameter count of 1.5 billion. Figure-\ref{fig:bm-gambar} summarizes the comparisons.

\section{Future Research}

Several research directions emerge from this foundational work. The extension to other Austronesian languages presents immediate opportunities for validation and refinement. Languages such as Bataknese, Javanese, Sundanese, and Malay share similar syllabic structures, potentially enabling the development of a unified tokenization framework for the language family. 

Adaptive tokenization mechanisms represent another promising avenue. Current frequency-based syllable selection could be enhanced through dynamic adjustment based on domain characteristics or task requirements. Information-theoretic optimization criteria could guide online adaptation of the syllable vocabulary to maintain optimal compression while accommodating domain-specific terminology.

Integration with contemporary language model architectures warrants systematic investigation. While evaluation employed GPT-2 architecture, modern developments in attention mechanisms, position encodings, and parameter-efficient training may interact differently with syllable-based tokenization. Systematic ablation studies across architectural variants would illuminate these interactions.

Theoretical analysis of compression bounds specific to syllable-based segmentation could provide principled guidance for vocabulary size selection and segmentation strategy optimization. Establishing connections between linguistic typology parameters and achievable compression rates would advance both theoretical understanding and practical implementation.

\section{Concluding Remarks}

This investigation introduces a linguistically-motivated tokenization approach for Indonesian language models, synthesizing insights from the Gasing Literacy Learning System with information-theoretic optimization principles. The empirical evaluation demonstrates that syllabic-agglutinative tokenization substantially outperforms conventional approaches across compression efficiency, token distribution balance, and vocabulary economy metrics.

The approach achieves Rényi efficiency values of 0.74 and 0.68 on Indonesian Wikipedia and folklore PDBI corpora respectively, while maintaining average token lengths exceeding those of both specialized and pretrained alternatives. These performance characteristics emerge from the fundamental alignment between syllable units and Indonesian morphophonological structure, enabling natural capture of linguistic regularities within the tokenization framework.

The theoretical analysis reveals how syllable-based segmentation effectively internalizes character-level dependencies, reducing the computational burden on downstream model components. By preserving morphological boundaries and respecting the agglutinative nature of Indonesian, the tokenization facilitates more efficient learning of compositional patterns essential for natural language understanding. This is demonstrated through an evaluative comparison of the prompts and responses produced by the TOBA-LM language model implementation against those of chatbot applications built on other popular, highly capable, multilingual language models.

The convergence of pedagogical insights from human literacy education with computational optimization principles suggests deeper connections between effective human language learning and machine processing strategies. As language technology increasingly serves diverse global communities, incorporating such linguistically-informed, culturally-grounded approaches becomes essential for developing equitable and efficient natural language processing systems.

This work establishes syllabic-agglutinative tokenization as a viable and advantageous approach for Indonesian language modeling, with broader implications for morphologically rich and underrepresented languages. The demonstrated benefits in compression efficiency, computational economy, and linguistic alignment provide a foundation for continued development of language technologies that respect and leverage the intrinsic structure of human languages.

\section*{Acknowledgement}

The author thank the Board Chairman of IT Del, H. E. Luhut Binsar Pandjaitan, for supporting the initiative of Gasing Method as well as support from fellows in Institut Teknologi Del.

\bibliographystyle{plain}

\begin{thebibliography}{99}

\bibitem{bostrom2020byte}
K. Bostrom and G. Durrett.
\newblock Byte pair encoding is suboptimal for language model pretraining.
\newblock arXiv:2004.03720v2, 2020.

\bibitem{cover2006elements}
T. Cover and J. Thomas.
\newblock {\em Elements of Information Theory}.
\newblock Wiley Series in Telecommunications and Signal Processing, 2006.

\bibitem{gage1994new}
P. Gage.
\newblock A new algorithm for data compression.
\newblock {\em The C Users Journal}, 12(2):23--38, 1994.

\bibitem{itzhak2021models}
I. Itzhak and O. Levy.
\newblock Models in a spelling bee: Language models implicitly learn the character composition of tokens.
\newblock arXiv:2108.11193, 2021.

\bibitem{kudo2018subword}
T. Kudo.
\newblock Subword regularization: Improving neural network translation models with multiple subword candidates.
\newblock In {\em Proceedings of the 56th Annual Meeting of the Association for Computational Linguistics (Volume 1: Long Papers)}, 2018.

\bibitem{libovicky2021why}
J. Libovick{\'y}, H. Schmid, and A. Fraser.
\newblock Why don't people use character-level machine translation?
\newblock arXiv:2110.08191, 2021.

\bibitem{makkuva2024attention}
A.~V. Makkuva et~al.
\newblock Attention with markov: A framework for principled analysis of transformers via markov chains.
\newblock arXiv:2402.04161, 2024.

\bibitem{petrov2023language}
A. Petrov et~al.
\newblock Language model tokenizers introduce unfairness between languages.
\newblock arXiv:2305.15425, 2023.

\bibitem{radford2019language}
A. Radford et~al.
\newblock Language models are unsupervised multitask learners.
\newblock OpenAI, 2019.

\bibitem{rajaraman2025toward}
N. Rajaraman, J. Jiao, and K. Ramchandran.
\newblock Toward a theory of tokenization in LLMs.
\newblock arXiv:2404.08335v2, 2025.

\bibitem{sennrich2016neural}
R. Sennrich, B. Haddow, and A. Birch.
\newblock Neural machine translation of rare words with subword units.
\newblock In {\em Proceedings of the 54th Annual Meeting of the Association for Computational Linguistics (Volume 1: Long Papers)}, 2016.

\bibitem{shannon1948mathematical}
C.~E. Shannon.
\newblock A mathematical theory of communication.
\newblock {\em The Bell System Technical Journal}, 27(3):379--423, 1948.

\bibitem{shannon1951prediction}
C.~E. Shannon.
\newblock Prediction and entropy of printed english.
\newblock {\em Bell System Technical Journal}, 30(1):50--64, 1951.

\bibitem{situngkir}
H. Situngkir.
\newblock Kode-kode Nusantara.
\newblock Expose-Mizan, 2016.

\bibitem{vaswani2017attention}
A. Vaswani et~al.
\newblock Attention is all you need.
\newblock arXiv:1706.03762v7, 2023.

\bibitem{zawani2008platform}
H. Zawani, D. Khanafiah, and H. Situngkir.
\newblock Platform komputasi untuk preservasi budaya tradisional secara partisipatif.
\newblock {\em BFI Working Paper Series}, 2008.

\bibitem{zouhar2023tokenization}
V. Zouhar et~al.
\newblock Tokenization and the noiseless channel.
\newblock arXiv:2306.16842v1, 2023.

\bibitem{zouhar2024formal}
V. Zouhar et~al.
\newblock A formal perspective on byte-pair encoding.
\newblock arXiv:2306.16837v3, 2024.

\end{thebibliography}

\end{document}